\documentclass[twocolumn]{aastex63} 

\usepackage{hyperref,enumerate}
\usepackage{xcolor}

\hypersetup{
    unicode=false,                  
    pdftoolbar=true,                
    pdfmenubar=true,                
    pdffitwindow=true,              
    pdfstartview={FitH},            
    pdftitle={SPLUSJ2104-0049},     
    pdfauthor={vplacco},            
    pdfsubject={Astronomy},         
    pdfcreator={dvipdf},            
    pdfproducer={dvipdf},           
    pdfkeywords={metal-poor stars}, 
    pdfnewwindow=true,              
    colorlinks=true,                
    linkcolor=red,                  
    citecolor=blue,                 
    filecolor=magenta,              
    urlcolor=cyan,                  
    breaklinks=true,
    linktocpage
}

\usepackage{hyperref,enumerate,numdef}

\newcommand{\kmsec}{\mbox{km~s$^{\rm -1}$}}
\newcommand{\eps}[1]{\ensuremath{\log\epsilon\,(\mathrm{#1})}}
\newcommand{\vv}{{\tablenotemark{\footnotesize{a}}}}
\newcommand{\xx}{{\tablenotemark{\footnotesize{b}}}}
\newcommand{\yy}{{\tablenotemark{\footnotesize{c}}}}
\newcommand{\ww}{{\tablenotemark{\footnotesize{d}}}}

\newcommand{\abund}[2]{\ensuremath{[\mathrm{#1}/\mathrm{#2}]}}
\newcommand{\cfe}{\abund{C}{Fe}}

\newcommand{\xfe}[1]{\abund{#1}{Fe}}

\newcommand{\metal}{\abund{Fe}{H}}

\newcommand{\teff}{\ensuremath{T_\mathrm{eff}}}
\newcommand{\logg}{\ensuremath{\log\,g}}
\newcommand{\Msun}{M_\odot}

\newcommand{\A}[1]{\ensuremath{A(\mathrm{#1})}}

\newcommand{\K}{\ensuremath{\mathrm{K}}}

\newcommand{\rave}{\object{SPLUS~J2104$-$0049}}
\newcommand{\ravel}{\object{SPLUS~J210428.01$-$004934.2}}

\newcommand{\splus}{S-PLUS}
\newcommand{\jplus}{J-PLUS}

\shortauthors{Placco et al.}

\begin{document}

\title{\ravel: An Ultra Metal-Poor Star Identified from Narrowband Photometry
\footnote{
Based on observations gathered with the $6.5\,$m Magellan
Telescopes located at Las Campanas Observatory, Chile.
Based on observations obtained at the international Gemini Observatory, a
program of NSF’s NOIRLab, which is managed by the Association of Universities
for Research in Astronomy (AURA) under a cooperative agreement with the
National Science Foundation on behalf of the Gemini Observatory
partnership: the National Science Foundation (United States), National
Research Council (Canada), Agencia Nacional de Investigaci\'{o}n y
Desarrollo (Chile), Ministerio de Ciencia, Tecnolog\'{i}a e Innovaci\'{o}n
(Argentina), Minist\'{e}rio da Ci\^{e}ncia, Tecnologia, Inova\c{c}\~{o}es e
Comunica\c{c}\~{o}es (Brazil), and Korea Astronomy and Space Science
Institute (Republic of Korea).  
}}

\author[0000-0003-4479-1265]{Vinicius M.\ Placco}
\affiliation{Community Science and Data Center/NSF’s NOIRLab, 950 N. Cherry Ave., Tucson, AZ 85719, USA}

\author[0000-0001-5107-8930]{Ian U.\ Roederer}
\affiliation{Department of Astronomy, University of Michigan, Ann Arbor, MI 48109, USA}
\affiliation{JINA Center for the Evolution of the Elements, USA}

\author{Young Sun Lee}
\affiliation{Department of Astronomy and Space Science, Chungnam National University, Daejeon 34134, South Korea}

\author{Felipe~Almeida-Fernandes}
\affiliation{Departamento de Astronomia, Instituto de Astronomia, Geof\'isica e
Ci\^encias Atmosf\'ericas da USP, Cidade \\ Universit\'aria, 05508-900, S\~ao
Paulo, SP, Brazil}

\author[0000-0001-7907-7884]{F\'abio R.~Herpich}
\affiliation{Departamento de Astronomia, Instituto de Astronomia, Geof\'isica e
Ci\^encias Atmosf\'ericas da USP, Cidade \\ Universit\'aria, 05508-900, S\~ao
Paulo, SP, Brazil}

\author[0000-0002-0537-4146]{H\'elio D. Perottoni}
\affiliation{Departamento de Astronomia, Instituto de Astronomia, Geof\'isica e
Ci\^encias Atmosf\'ericas da USP, Cidade \\ Universit\'aria, 05508-900, S\~ao
Paulo, SP, Brazil}

\author[0000-0002-4064-7234]{William Schoenell}
\affiliation{GMTO Corporation 465 N. Halstead Street, Suite 250 Pasadena, CA 91107, USA}

\author{Tiago Ribeiro}
\affiliation{Rubin Observatory Project Office, 950 N. Cherry Ave., Tucson, AZ 85719, USA }

\author{Antonio Kanaan}
\affiliation{Departamento de F\'isica, Universidade Federal de Santa Catarina, Florian\'opolis, SC 88040-900, Brazil}

\correspondingauthor{Vinicius M.\ Placco}
\email{vinicius.placco@noirlab.edu}

\begin{abstract}

We report on the discovery of \ravel, an ultra metal-poor (UMP) star first
identified from the narrow-band photometry of the Southern Photometric Local
Universe Survey (S-PLUS) Data Release 1, in the SDSS Stripe 82 region. Follow-up
medium- and high-resolution spectroscopy (with Gemini South and Magellan-Clay,
respectively) confirmed the effectiveness of the search for low-metallicity
stars using the S-PLUS narrow-band photometry. At \metal=$-4.03$, \rave\ has the lowest
{\emph{detected}} carbon abundance, A(C)=$+$4.34, when compared to the 34
previously known UMP stars in the literature, which is an important constraint
on its stellar progenitor and also on stellar evolution models at the lowest
metallicities. Based on its chemical abundance pattern, we speculate that
\rave\ could be a bona-fide second-generation star, formed from a gas cloud
polluted by a single metal-free $\sim 30 \Msun$ star.
This discovery opens the possibility of finding additional UMP stars directly
from narrow-band photometric surveys, a potentially powerful method to help
complete the inventory of such peculiar objects in our Galaxy.

\end{abstract}


\section{Introduction}
\label{intro}

Is there any observational evidence that the first generation of stars
born in the universe (Population III; hereafter Pop III) had an initial mass 
function (IMF) that allowed the formation of low-mass ($M \leq 1.0 \Msun$)
objects?
Cosmological simulations indicate that the Pop III IMF can extend to such low
masses \citep{stacy2016}.
However, as of today, no metal-free stars have been found. Even the most chemically
pristine star ever observed \citep[SMSS~J031300.36$-$670839.3;][]{keller2014}
has lithium, carbon, oxygen, magnesium, and calcium detected in its atmosphere.
Based on current theoretical work, molecular hydrogen cooling allows the
formation of minihalos of $10^6 \Msun$  as early as $z \approx 20-30$, which
will fragment and form predominantly massive ($M>10\Msun$) stars
\citep[][]{bromm2013}.
Then, with the first chemical elements heavier than He introduced in the
interstellar medium by the evolution of these massive objects, the formation of
low-mass objects would be facilitated by additional cooling channels, such as
dust and metal lines \citep[in particular \ion{C}{2} and
\ion{O}{1};][]{dopcke2013}. 
Alternatively, \citet{schlaufman2018} found evidence implying that it is
possible to have surviving (present day) solar-mass stars that were secondaries
around massive Pop III stars ($10 \leq M/\Msun \leq 100$), and were formed via
disk fragmentation.

Ultra Metal-Poor (UMP; [Fe/H]\footnote{\abund{A}{B} = $\log(N_X/{}N_Y)_{\star} -
\log(N_X/{}N_Y) _{\odot}$, where $N$ is the number density of elements
$X$ and $Y$ in the star ($\star$) and the Sun ($\odot$).} $<
-4.0$) stars \citep{beers2005}, while still members of the second generation, can provide an
observational benchmark as to whether such low-mass metal-free stars exist.
According to \citet{hartwig2015}, in order to rule out (at a 99\% confidence
level) the existence of a low-mass metal-free star, $\sim 2 \times 10^7$ halo
stars should be observed and have their \metal\ determined. That translates
into roughly several hundred observed UMP stars, although only
34\footnote{High-resolution ($R \geq 20,000$) spectroscopy is required to
derive \metal\ and classify a star as an UMP.} have been found to date
\citep{saga2008,jinabase}.


One technique to select suitable UMP candidates for spectroscopic follow-up is
through photometric metallicities.
The first effort of estimating the
metallicity from the Sloan Digital Sky
Survey \citep[SDSS;][]{york2000} $u-g$ and $g-r$ colors was published
by \citet{ivezic2008}. The authors were able to determine
\metal\ for over 2 million F/G stars in the Milky Way with uncertainties of 
0.2~dex or better for $-2.0 \leq $\metal$ \leq -0.5$.
However, due to the intrinsic broadness of the $u$ filter, which carries
most of the metallicity information, the uncertainties increase considerably for
\metal$\leq -2.0$. 

%
The $u$ and $v$ filters from SkyMapper provide extra discriminating power
due to their ability to break the degeneracy between surface gravity and
metallicity. From its Data Release 1 \citep[DR1;][]{wolf2018},
photometric atmospheric parameters were determined with a precision better 
than $\sim 0.2$~dex for \metal$\geq -2.0$ \citep{casagrande2019}.
%
%
Another recent effort to search for low-metallicity stars in
the Milky-Way is the Pristine Survey \citep{starkenburg2017}, which employs
narrow-band photometry on the metallicity sensitive \ion{Ca}{2} K
absorption feature, in addition to SDSS $g$ and $i$ filters. The
$\sim 100$\,{\AA}-wide narrow-band filter 
is able to predict metallicities down to \metal$\sim -3.0$. A spectroscopic follow-up campaign shows
that, out of the 1007 stars observed, $\sim 70\%$ have \metal$< -2.0$ and $\sim 9\%$
have \metal$< -3.0$ \citep{aguado2019}. 

\begin{figure*}[!ht]
\epsscale{1.11}
\plotone{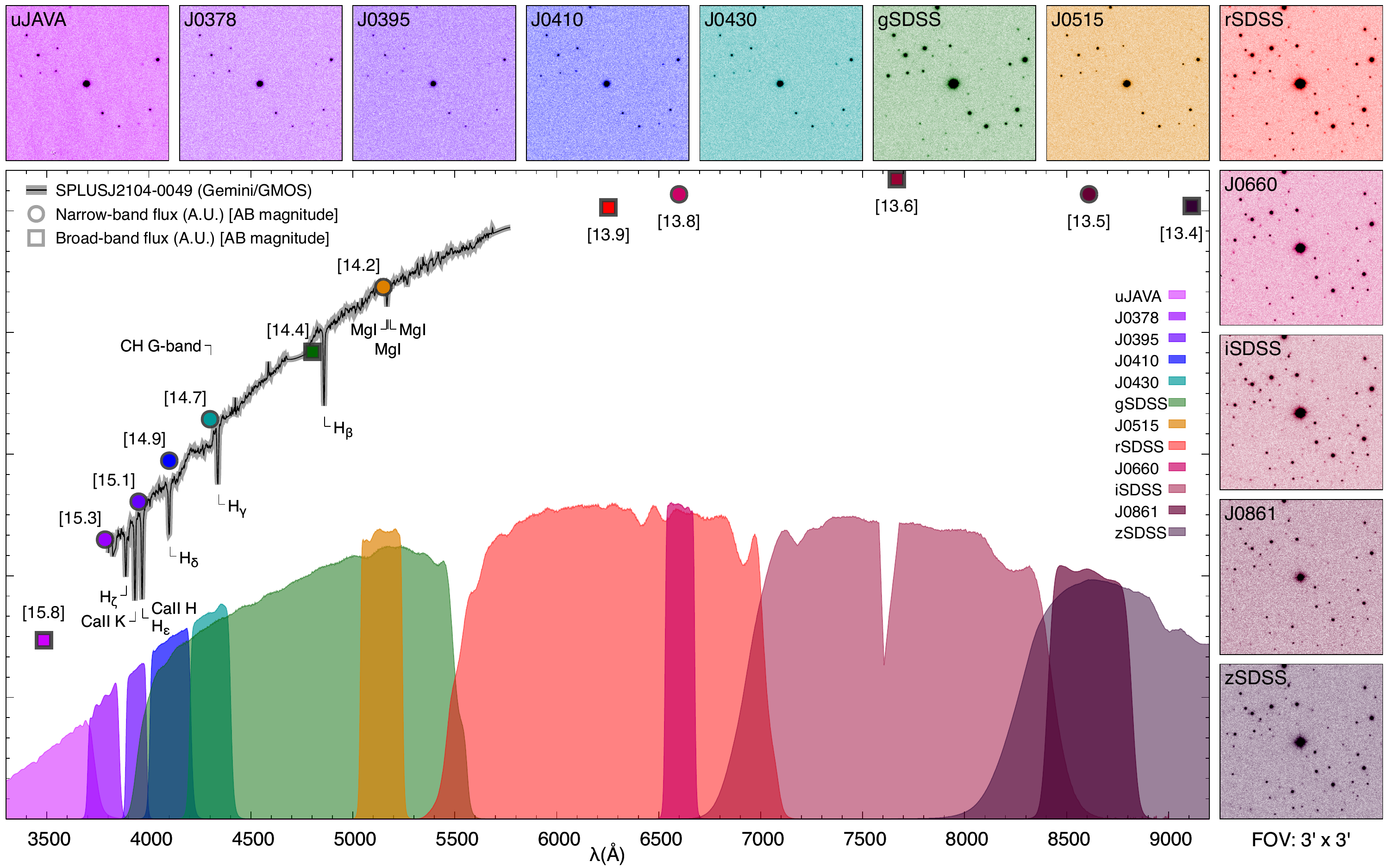}
\caption{Outside panels: 12-band S-PLUS images for \protect\rave, retrieved
from the Astro Data Lab. The field of view is 3'~$\times$~3', with the north
direction up and east to the left. The color of each image is based on the
central wavelength of the Javalambre filters, which are named on the top left
part of each image.  Main panel: transmission curves measured for the set of 12
Javalambre filters, which include the effect of the entire system (sky,
mirrors, lenses, and CCD). Also shown are the Gemini/GMOS spectrum (black solid
line), the fluxes in the narrow-band (filled circles) and broad-band (filled
squares) filters -- calculated from the AB magnitudes (in square brackets).}
\label{splus} 
\end{figure*}

The next generation of narrow-band photometric surveys is already underway,
building (and improving) upon the successes described above. Two such efforts
are the
Javalambre Photometric Local Universe Survey \citep[\jplus;][]{cenarro2019} and
the Southern Photometric Local Universe Survey
\citep[\splus;][]{mendesdeoliveira2019}.
Both surveys have identical fully-robotic telescopes with 0.83~m mirrors and
2.0~deg$^2$ field of view, performing precision multiple-filter optical 
photometry (3500\,{\AA} to 10,000\,{\AA}) with a set of 12 broad- and
narrow-band filters, consisting of four SDSS-like ($g$SDSS, $r$SDSS, $i$SDSS, $z$SDSS), one modified
SDSS $u$, and seven narrow-band (100-400\,{\AA} FWHM) filters. 
Figure~\ref{splus} shows the Javalambre photometric system.
These filters, by virtue of their restricted bandpasses, have a
much higher sensitivity for the determination of stellar atmospheric parameters
and selected chemical abundances.  In the first attempt to determine
metallicities from \jplus\ photometry, \cite{whitten2019} were able to
successfully reproduce spectroscopic values down to \metal$\sim -3.5$ with a
standard deviation of the residuals $\sigma \sim 0.25$~dex. More recently,
\citet{whitten2021} were able to calculate photometric \teff, \metal, and, for
the first time, carbon abundances for over 700,000 stars in the \splus\ DR2 
with similar precision.

We report the discovery of \ravel\ (hereafter \rave), an UMP
star selected from its narrow-band S-PLUS photometry and confirmed by medium-
and high-resolution spectroscopy. These proof-of-concept observations are part
of an ongoing effort to spectroscopically confirm low-metallicity
candidates identified from narrow-band photometry.


\begin{deluxetable*}{lllll}
\tablecaption{Observational Data for \protect\ravel \label{starinfo}}
\tablewidth{0pt}
\tabletypesize{\scriptsize}
\tabletypesize{\small}
\tablehead{
\colhead{Quantity} &
\colhead{Symbol} &
\colhead{Value} &
\colhead{Units} &
\colhead{Reference}}
\startdata
Right ascension           & $\alpha$ (J2000)    & 21:04:28.01            & hh:mm:ss.ss   & \citet{gaia2020}             \\ 
Declination               & $\delta$ (J2000)    & $-$00:49:34.2          & dd:mm:ss.s    & \citet{gaia2020}             \\ 
Galactic longitude        & $\ell$              & 48.7700                & degrees       & \citet{gaia2020}             \\ 
Galactic latitude         & $b$                 & $-$29.6429             & degrees       & \citet{gaia2020}             \\ 
Gaia EDR3 Name            &                     & 2689845933385992064    &               & \citet{gaia2020}             \\ 
Parallax                  & $\varpi$            & 0.1619 $\pm$ 0.0245    & mas           & \citet{gaia2020}             \\ 
Inverse parallax distance & $1/\varpi$          & 4.92$^{+0.67}_{-0.53}$ & kpc           & This study\vv                \\ 
Proper motion ($\alpha$)  & PMRA                & 14.976 $\pm$ 0.027     & mas yr$^{-1}$ & \citet{gaia2020}             \\ 
Proper motion ($\delta$)  & PMDec               & $-$8.260 $\pm$ 0.017   & mas yr$^{-1}$ & \citet{gaia2020}             \\ 
Mass                      & $M$                 & 0.80 $\pm$ 0.15        & $M_{\odot}$   & Assumed                      \\ 
$B$ magnitude             & $B$                 & 14.978 $\pm$ 0.051     & mag           & \citet{henden2014}           \\ 
$V$ magnitude             & $V$                 & 14.245 $\pm$ 0.095     & mag           & \citet{henden2014}           \\ 
$J$ magnitude             & $J$                 & 12.546 $\pm$ 0.023     & mag           & \citet{skrutskie2006}        \\ 
$H$ magnitude             & $H$                 & 12.052 $\pm$ 0.024     & mag           & \citet{skrutskie2006}        \\ 
$K$ magnitude             & $K$                 & 11.968 $\pm$ 0.028     & mag           & \citet{skrutskie2006}        \\ 
Color excess              & $E(B-V)$            & 0.0557 $\pm$ 0.0019    & mag           & \citet{schlafly2011}         \\ 
Bolometric correction     & BC$_V$              & $-$0.54 $\pm$ 0.08     & mag           & \citet{casagrande2014}       \\ 
Radial velocity           & RV                  & $-$110.3 $\pm$ 0.5     & \kmsec        & Magellan (MJD: 59166.0389)   \\ 
Effective Temperature     & \teff               & 5045$^{+210}_{-95}$    & K             & \citet{gaia2020}             \\ 
                          &                     & 5044 $\pm$ 150         & K             & This study (Gemini)          \\ 
                          &                     & 4812 $\pm$ 55          & K             & This study (Magellan)        \\ 
Log of surface gravity    & \logg               & 2.66 $\pm$ 0.20        & (cgs)         & This study (Gemini)          \\ 
                          &                     & 1.95 $\pm$ 0.17        & (cgs)         & This study (Magellan)        \\ 
Microturbulent velocity   & $\xi$               & 1.60 $\pm$ 0.20        & \kmsec        & This study (Magellan)        \\ 
Metallicity               & \metal              & $-$4.22 $\pm$ 0.20     & dex           & This study (Gemini)          \\ 
                          &                     & $-$4.19 $\pm$ 0.06     & dex           & This study LTE (Magellan)    \\ 
                          &                     & $-$4.03 $\pm$ 0.10     & dex           & This study NLTE (Magellan)   \\ 
\enddata
\tablenotetext{a}{Using $\varpi_{\rm zp} = -0.0414$ mas from \citet{lindegren2020}.}
\end{deluxetable*}

\begin{figure*}[!ht]
\epsscale{1.04}
\plotone{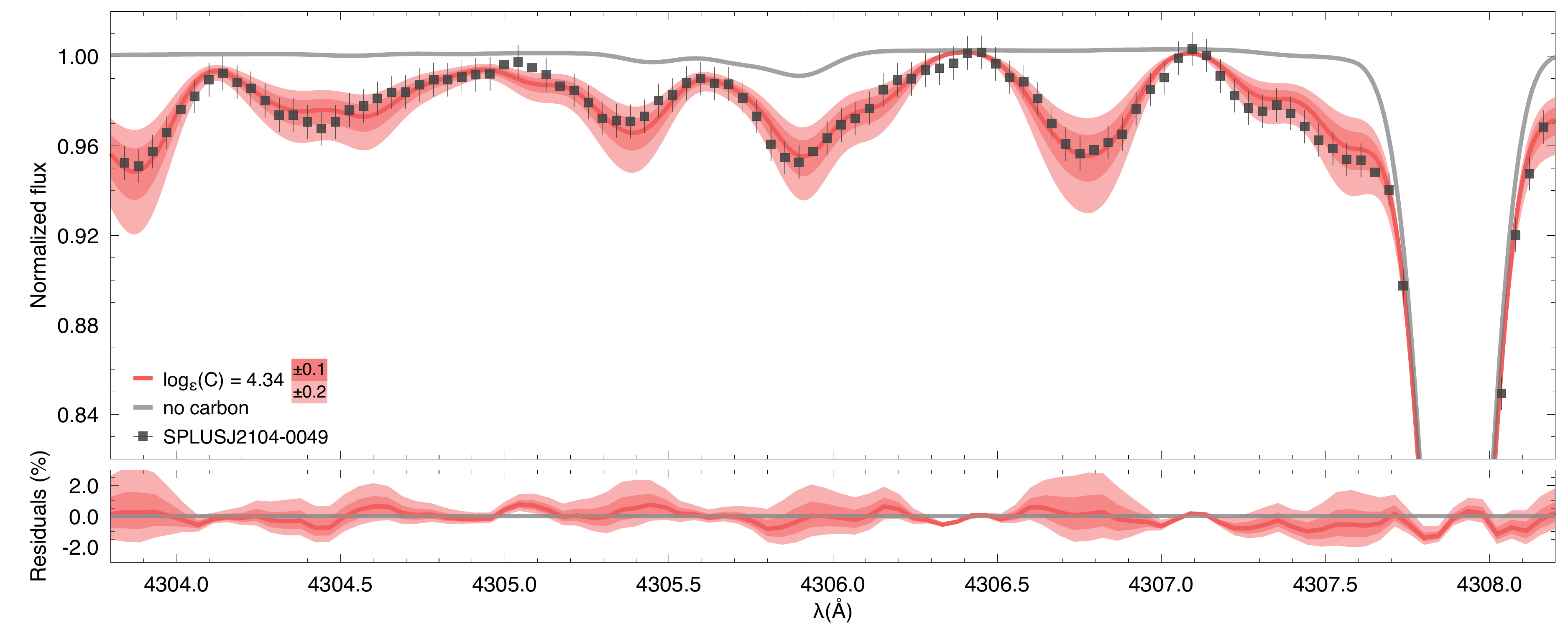}
\epsscale{0.5}
\plotone{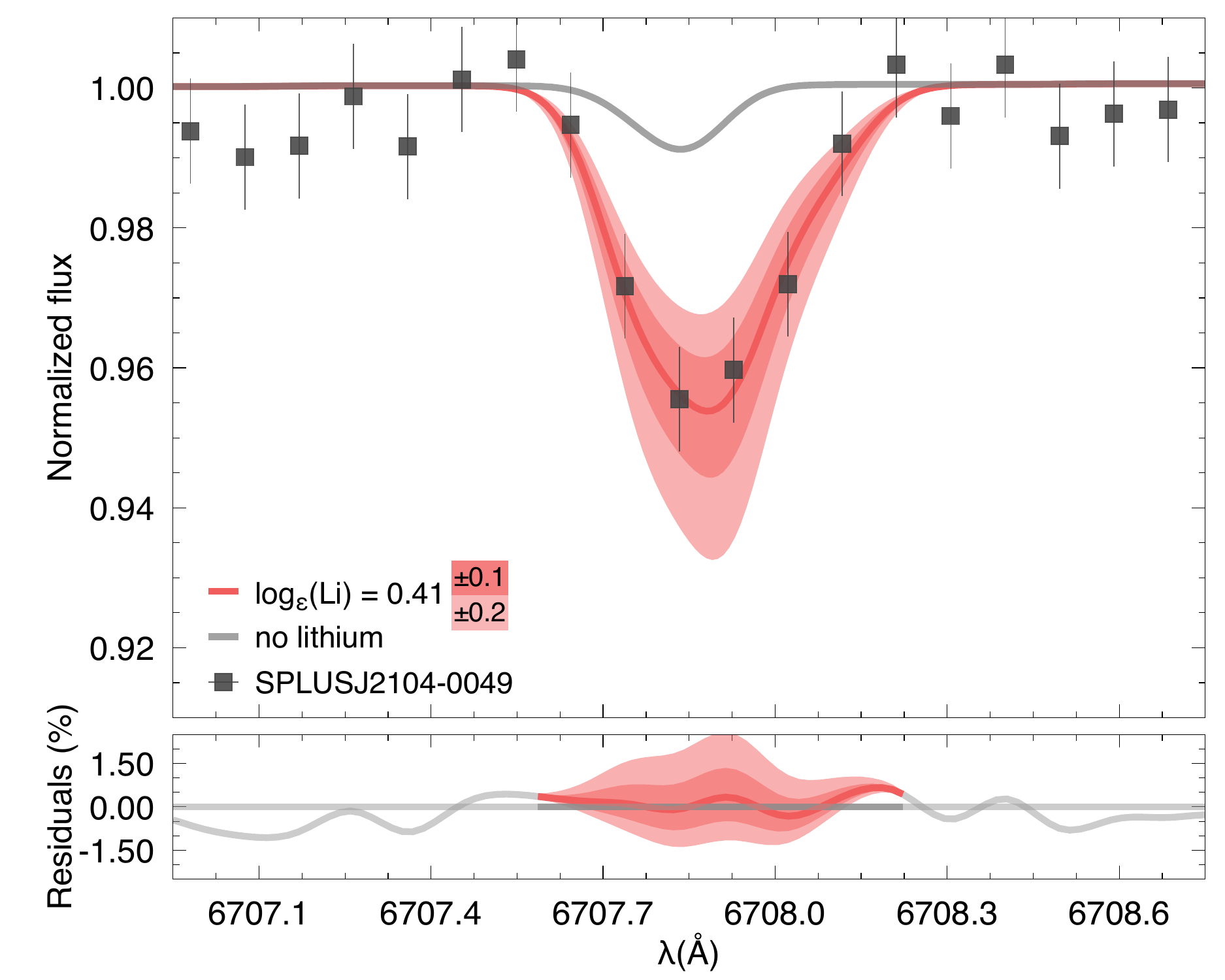}
\epsscale{0.5}
\plotone{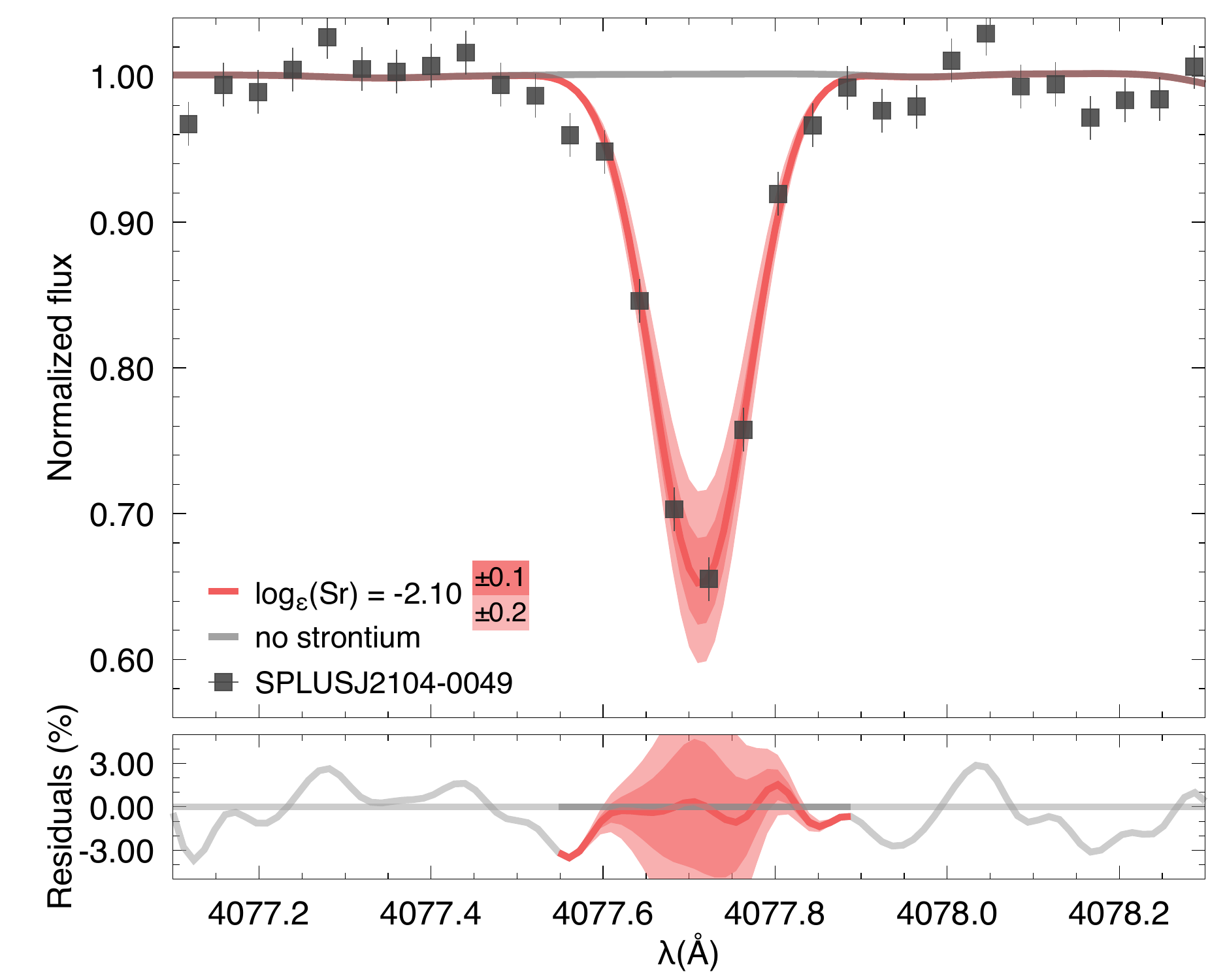}
\caption{Spectral syntheses for the determination of carbon (upper panel),
lithium (lower left panel), and strontium (lower right panel) abundances. The
top panel of each plot shows the best-fit syntheses (red lines) and
uncertainties ($\pm 0.1$ and $\pm 0.2$~dex - shaded regions) compared to the
observed spectra (points). Also shown are syntheses after removing the
contributions from specific elements (gray lines).The bottom panels show the \%
residuals between the observed spectra and the syntheses.}
\label{synthesis}
\end{figure*}


\section{Target Selection and Observations}
\label{secobs}

\subsection{Narrow-band Photometry}

The 12-band photometric data for \rave\ was obtained during the first S-PLUS
observing campaign (Data Release 1 - DR1\footnote{The photometry and
images are publicly available at the NSF's NOIRLab Astro Data Lab:
\href{https://datalab.noirlab.edu/splus/}{https://datalab.noirlab.edu/splus/}}) on the
Stripe 82, which is a $\sim$336~deg$^2$ equatorial field that was first imaged
several times by SDSS.

\rave\ is part of a larger sample of metal-poor star candidates selected based
on their position on a color-color diagram constructed using
metallicity-sensitive magnitudes, such as J0395 and J0515. 
Details on the target selection, its effectiveness in identifying chemically
peculiar stars, and the spectroscopic follow-up are the subject of a forthcoming
paper (Placco et al., in preparation).  Table~\ref{starinfo} summarizes
information about \rave. 
%
Figure~\ref{splus} shows the 12-band S-PLUS images for \rave.  Also shown on
the main panel are the transmission curves measured for the set of 12 filters,
as well as the AB magnitudes (in square brackets).

\subsection{Medium-resolution Spectroscopy}

The first spectroscopic follow-up of \rave\ was conducted with the Gemini South
Telescope on 2019 May 17, as part of the poor weather program GS-2019A-Q-408.
The GMOS-S
instrument was used with the B600~l~mm$^{\rm{-1}}$ grating (G5323) and a
1$\farcs$0 slit with $2 \times 2$ binning, resulting in a wavelength coverage in
the range [3600:5800]\,{\AA} at resolving power $R \sim 2,000$.  The 1,200~s 
exposure resulted in a signal-to-noise ratio of S/N\,$ \sim 100$ per
pixel at the \ion{Ca}{2}~K line (3933.3\,{\AA}).  Calibration frames included
arc-lamp exposures, bias frames, and quartz flats. All tasks related to spectral
reduction, extraction, and wavelength calibration were performed using the
Gemini
IRAF\footnote{\href{https://www.gemini.edu/observing/phase-iii/understanding-and-processing-data/Data-Processing-Software}
{https://www.gemini.edu/observing/phase-iii/understanding-and-processing-data/Data-Processing-Software}.}
standard routines.

The central panel of Figure~\ref{splus} shows the Gemini/GMOS data, scaled
in flux by convolving the normalized spectrum with a blackbody curve at
\teff=4800~K. Prominent absorption features are identified.

\subsection{High-resolution Spectroscopy}

The final confirmation step for \rave\ was the high-resolution spectroscopy,
obtained on 2020 November 13 using the MIKE spectrograph mounted on the
6.5m Magellan-Clay Telescope at Las Campanas Observatory.  The observing setup
included a $0\farcs7$ slit with $2\times2$ on-chip binning, yielding a resolving
power of $R\sim37,000$ ($\lambda<5000\,${\AA}) and $R\sim30,000$ ($\lambda>5000\,${\AA}). The S/N
is $\sim40$ per pixel at $3900\,${\AA} and $\sim 120$ at $5200\,${\AA} after
3,200s of exposure time. The MIKE spectrum covers most of the optical
wavelength regime ($\sim3300-9000\,${\AA}), making it suitable for chemical
abundance determinations.  The blue and red MIKE spectra were reduced using the
routines described in
\citet{kelson2003}\footnote{\href{http://code.obs.carnegiescience.edu/python}
{http://code.obs.carnegiescience.edu/python}}.

\section{Stellar Atmospheric Parameters}
\label{secatm}

Stellar atmospheric parameters (\teff, \logg, and \metal) were calculated from
the Gemini/GMOS spectrum using the n-SSPP \citep{beers2014}, which is
adapted from the SEGUE Stellar Parameter Pipeline
\citep[SSPP;][]{lee2008a}.  These parameters were used to
select \rave\ as a candidate for high-resolution spectroscopic follow-up. 
Table~\ref{starinfo} lists \teff, \logg, and \metal\ derived from the Gemini
spectrum.

The stellar parameters for the high-resolution data were determined from a
combination of photometry, the Gaia parallax
\citep{gaia2020}, and the MIKE spectrum.
The effective temperature for \rave\ was calculated from the
metallicity-dependent color-\teff\ relations by \citet{casagrande2010}, adopting
\metal=$-4.0\pm0.2$. We used the same procedure outlined in
\citet{roederer2018}, drawing 10$^5$ samples for magnitudes, reddening, and
metallicity. The final \teff=$4812\pm55$~K is the weighted mean of the median
temperatures for each input color ($B-V$, $V-J$, $V-H$, $V-K$, and $J-K$).
The surface gravity was calculated using Equation~1 in \citet{roederer2018},
drawing 10$^5$ samples from the input parameters listed in Table~\ref{starinfo}.
The final \logg=$1.95\pm0.17$ is taken as the median of those calculations with the
uncertainty given by their standard deviation.

The equivalent widths were obtained by fitting Gaussian profiles to the observed
absorption features. With \teff\ and \logg\ determined above, the
\ion{Fe}{1} abundances were determined spectroscopically, using the latest
version of the
MOOG\footnote{\href{https://github.com/alexji/moog17scat}{https://github.com/alexji/moog17scat}}
code \citep{sneden1973}, employing one-dimensional plane-parallel
model atmospheres with no overshooting \citep{castelli2004}, assuming local
thermodynamic equilibrium (LTE). 
No reliable \ion{Fe}{2} features were found in the \rave\ MIKE spectrum. The
microturbulent velocity was determined by minimizing the trend between the
abundances of individual \ion{Fe}{1} absorption features and their reduced
equivalent width.
The mean LTE abundance from 51 \ion{Fe}{1} lines is \metal=$-4.19\pm0.06$. For
19 of those absorption features, we were able to determine non-LTE (NLTE)
abundances using version 1.0 of the
{\texttt{INSPECT}}\footnote{\href{http://www.inspect-stars.com/}{http://www.inspect-stars.com/}}
database \citep{lind2012,bergemann2012b}.
The average difference between the LTE and NLTE abundances is
$\Delta$NLTE=$+0.16\pm0.03$ and the adopted \rave\ metallicity for the remainder
of this work is \metal=$-4.03\pm0.10$. 
Table~\ref{starinfo} lists the final atmospheric parameters for \rave, which
will be used for the abundance analysis.

\section{Chemical Abundances}
\label{absec}

Elemental-abundance ratios, \xfe{X}, were calculated adopting the Solar
photospheric abundances from \citet{asplund2009}.  The average measurements for
$18$ elements, derived from the Magellan/MIKE spectrum, are listed in
Table~\ref{abund}.  The $\sigma$ values are the standard error of the mean.
For $\sigma$ values below 0.10~dex we set a standard fixed uncertainty of
0.10~dex.  
For elements with only one detected absorption feature, the uncertainty is
determined from the spectral synthesis (cf. Figure~\ref{synthesis}).
The last column shows which elements had their abundances calculated
via equivalent-width analysis (\texttt{eqw}) or spectral synthesis
(\texttt{syn}). The atomic and molecular line lists were generated by the
\texttt{linemake}\footnote{\href{https://github.com/vmplacco/linemake}{https://github.com/vmplacco/linemake}}
code \citep{placco2021}. Individual references are given in their README file.
We have determined NLTE abundance corrections for three elements besides
\ion{Fe}{1}: \ion{Al}{1}, \ion{Cr}{1}, and \ion{Mn}{1}. The values and
references are given in Table~\ref{abund}.

Overall, \rave\ has the chemical abundance pattern of a ``typical'' UMP star
(apart from carbon -- see discussion in Section~\ref{origin}). The
lithium abundance is consistent with its evolutionary stage and the
light-element abundance ratios \xfe{X} (from Na to Zn) are in agreement with
general trends found in the literature at this metallicity regime \citep{jinabase}. The same
applies to the low abundance ratios found for the heavy elements Sr and Ba.
The top panel of Figure~\ref{synthesis} shows the spectral synthesis of the CH
$G$-band at $\lambda$4304\,{\AA} for \rave. 
The lower panels show the same procedure for the Li~\textsc{i} $\lambda$6707\,{\AA} and Sr~\textsc{ii}
$\lambda$4077\,{\AA} absorption features.
Even though \rave\ is on the red-giant branch, there is no carbon depletion due
to CN processing, which is a result of the combination of its low metallicity and
low carbon abundance \citep[cf. Figure~9 in][]{placco2014c}.

\begin{deluxetable}{@{}lcrrrcrr@{}}
\tabletypesize{\small}
\tabletypesize{\footnotesize}
\tablewidth{0pc}
\tablecaption{Abundances for Individual Species \label{abund}}
\tablehead{
\colhead{Ion}                         & 
\colhead{$\log\epsilon_{\odot}$\,(X)} & 
\colhead{$\log\epsilon$\,(X)}         & 
\colhead{$\mbox{[X/H]}$}              & 
\colhead{$\mbox{[X/Fe]}$}             & 
\colhead{$\sigma$}                    & 
\colhead{$N$}                         &
\colhead{}}
\startdata
\ion{Li}{1}    & 1.05 &    0.41 & \nodata & \nodata & 0.15 &  1 & {\texttt{syn}} \\ 
\ion{C}{0}     & 8.43 &    4.34 & $-$4.09 & $-$0.06 & 0.15 &  3 & {\texttt{syn}} \\ 
\ion{Na}{1}    & 6.24 &    1.98 & $-$4.26 & $-$0.23 & 0.10 &  2 & {\texttt{eqw}} \\ 
\ion{Mg}{1}    & 7.60 &    3.94 & $-$3.66 & $+$0.37 & 0.10 &  4 & {\texttt{eqw}} \\ 
\ion{Al}{1}\vv & 6.45 &    2.37 & $-$4.08 & $-$0.05 & 0.15 &  2 & {\texttt{syn}} \\ 
\ion{Si}{1}    & 7.51 &    4.07 & $-$3.44 & $+$0.59 & 0.15 &  1 & {\texttt{syn}} \\ 
\ion{Ca}{1}    & 6.34 &    2.63 & $-$3.71 & $+$0.32 & 0.10 &  2 & {\texttt{syn}} \\ 
\ion{Sc}{2}    & 3.15 & $-$0.65 & $-$3.80 & $+$0.23 & 0.10 &  5 & {\texttt{eqw}} \\ 
\ion{Ti}{2}    & 4.95 &    1.22 & $-$3.73 & $+$0.30 & 0.10 &  9 & {\texttt{eqw}} \\ 
\ion{V}{2}     & 3.93 &    0.39 & $-$3.54 & $+$0.49 & 0.20 &  1 & {\texttt{syn}} \\ 
\ion{Cr}{1}\xx & 5.64 &    1.35 & $-$4.29 & $-$0.26 & 0.10 &  2 & {\texttt{eqw}} \\ 
\ion{Mn}{1}\yy & 5.43 &    0.75 & $-$4.68 & $-$0.65 & 0.10 &  2 & {\texttt{syn}} \\ 
\ion{Fe}{1}\ww & 7.50 &    3.47 & $-$4.03 &    0.00 & 0.10 & 19 & {\texttt{eqw}} \\ 
\ion{Co}{1}    & 4.99 &    1.15 & $-$3.84 & $+$0.19 & 0.10 &  3 & {\texttt{eqw}} \\ 
\ion{Ni}{1}    & 6.22 &    2.22 & $-$4.00 & $+$0.03 & 0.10 &  3 & {\texttt{eqw}} \\ 
\ion{Zn}{1}    & 4.56 &    1.17 & $-$3.39 & $+$0.64 & 0.20 &  1 & {\texttt{syn}} \\ 
\ion{Sr}{2}    & 2.87 & $-$2.07 & $-$4.94 & $-$0.91 & 0.15 &  2 & {\texttt{syn}} \\ 
\ion{Ba}{2}    & 2.18 & $-$3.06 & $-$5.24 & $-$1.21 & 0.20 &  2 & {\texttt{syn}} \\ 
\enddata
\tablenotetext{a}{$\Delta$NLTE=$+0.60$ \citep{nordlander2017b}}
\tablenotetext{b}{$\Delta$NLTE=$+0.35$ \citep[based on the empirical corrections of][]{roederer2014}}
\tablenotetext{c}{$\Delta$NLTE=$+0.60$ \citep{bergemann2008}}
\tablenotetext{d}{$\Delta$NLTE=$+0.16$ \citep{bergemann2012b,lind2012}}
\end{deluxetable}

\begin{figure*}[!ht]
\epsscale{1.075}
\plotone{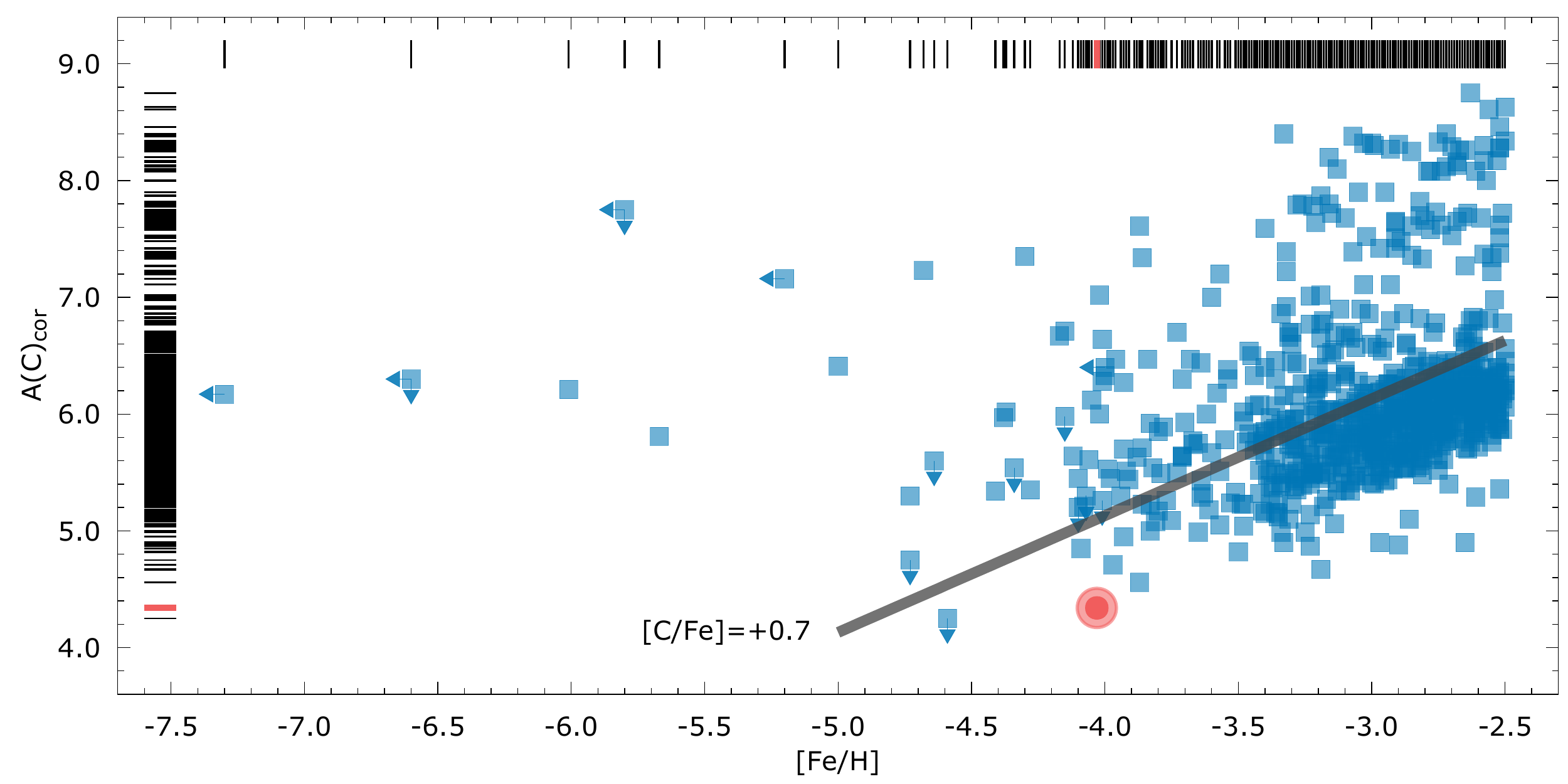}
\plotone{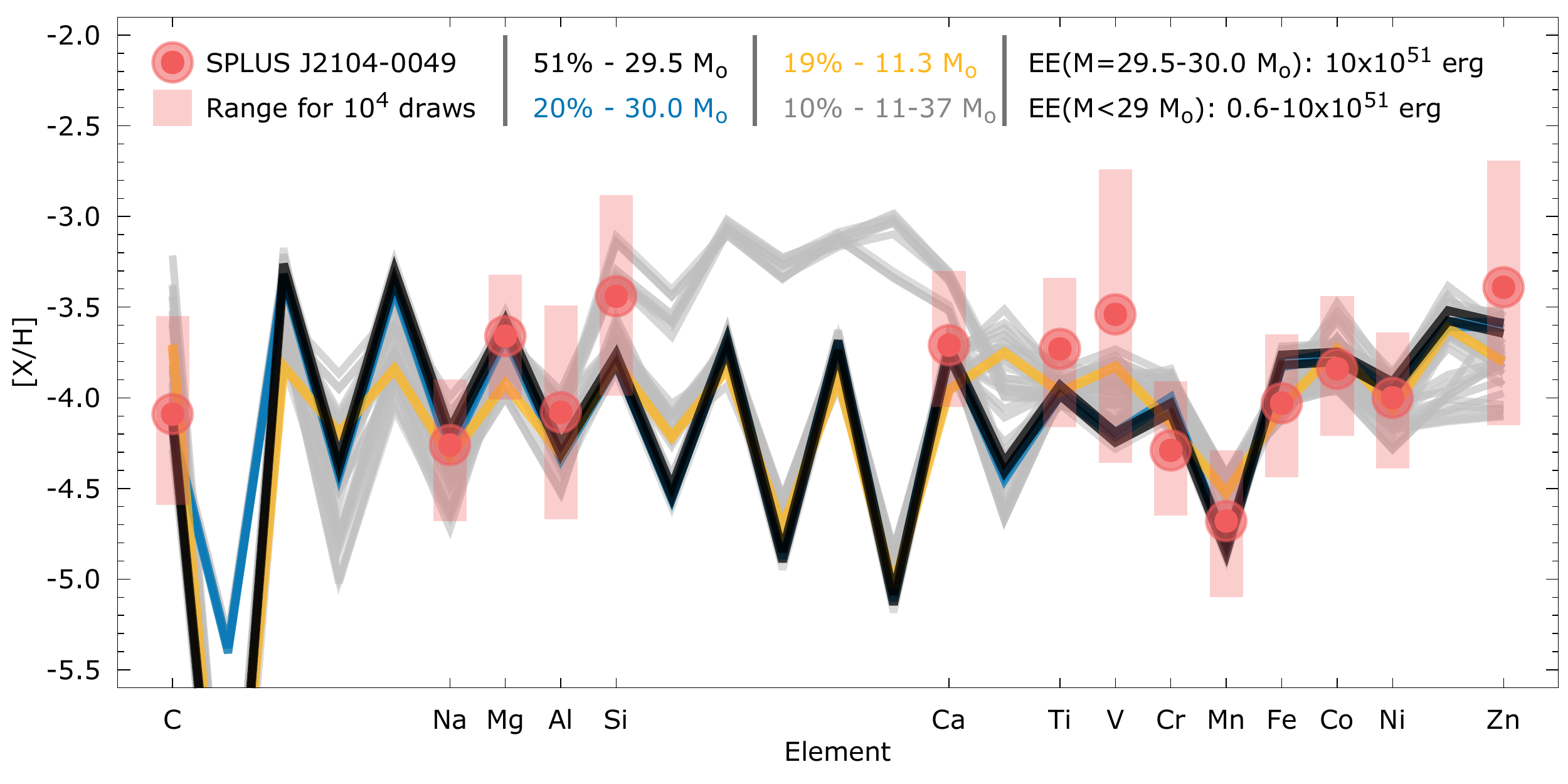}
\caption{Upper panel: Carbon abundances -- A(C) -- as a function of the metallicity --
\protect\metal\ -- for \protect\rave\ (filled circle) and the literature
compilations JINAbase \citep{jinabase} and SAGA \citep{saga2008} (filled
squares). The solid line represents the current criteria for CEMP stars
(\protect\cfe=$+0.7$).  Upper limits are only shown for \metal$\leq -4.0$.
Stripe-density profiles are also shown, with the values
for \protect\rave\ highlighted. An interactive version of this panel is available at 
\href{http://vmplacco.github.io/files/acfeh.html}{http://vmplacco.github.io/files/acfeh.html}.
Lower panel: Best model fits for \rave. The
solid lines show the theoretical predictions from the \citet{heger2010}
\texttt{znuc2012.S4} models, color coded by mass and occurence fraction within
the 10$^4$ simulations. The explosion energies (EE) are also listed.
The solid circles are the measured abundances for
\protect\rave\ and the shaded areas mark the range of simulated abundances.}
\label{acfeh}
\end{figure*}

\vspace{0.5cm}

\section{Possible Origins for \protect\rave}
\label{origin}

The current working hypothesis in stellar archaeology is that UMP stars are
{\emph{bona fide}} second-generation objects chemically enriched by a single
Pop. III supernova (``mono-enriched''); thus their chemical abundance
pattern is a direct result of the composition of the parent gas cloud. Below we
present possible formation pathways and stellar progenitors that could account
for the existence of \rave\ and its low carbon.

\rave\ is the 35th UMP star identified to date
\citep{saga2008,jinabase}\footnote{The SAGA database was last updated on
2020-11-09.}.  Among these, only three {\emph{are not}} classified as
Carbon-Enhanced Metal-Poor (CEMP; \cfe$\geq+0.7$, \citealt{aoki2007}):
CD$-$38$^\circ$245 \citep[\cfe$<-0.33$]{spite2005}, CS~22963$-$004
\citep[\cfe=$+0.40$]{lai2008}, and now \rave\ (\cfe=$-0.06$), with the
lowest $A$(C)\footnote{$A$(C) = $\log(N_C/{}N_H) + 12$.} value ever
{\emph{detected}} in the \metal$<-4.0$ regime. The upper panel of
Figure~\ref{acfeh} shows the $A$(C)$_{\rm cor}$\footnote{The observed $A$(C)
values have been corrected following the prescriptions found in
\citet{placco2014c}.} vs. \metal\ distribution for stars in the literature with
\metal$<-2.5$ (blue filled squares) compared to \rave\ (red filled circle). Also
shown are stripe-density profiles and the line defining the CEMP criteria. 
Based on these data (and with the addition of \rave), the CEMP fraction among
UMP stars is $91^{+6}_{-14}\%$\footnote{Uncertainties in the fractions are the
Wilson score confidence intervals.} (32/35 - including $A$(C)
upper limits) and $92^{+6}_{-17}\%$ (23/25 - excluding upper limits). These are
larger, nonetheless consistent, with the 81\% fraction calculated by
\citet{placco2014c}. 

The low carbon abundance in \rave\ helps constrain the main cooling channel
that allowed its parent gas cloud to fragment. According to \citet{chiaki2017},
there is insufficient cooling from carbon dust grains for $A$(C)$\lesssim5.8$,
so the most efficient way to induce cloud fragmentation would be by silicate
dust cooling. In fact, \rave\ resides in the ``silicate dominant'' regime
in the $A$(C)-\metal\ diagram \citep[c.f. Figure 2 of][]{chiaki2017}.
An additional diagnostic to assess whether a star is ``mono-enriched'' is
through its \abund{Mg}{C} abundance ratio \citep{hartwig2018}. The
low-metallicity of \rave, coupled with its \abund{Mg}{C}=$+0.43$, places it
well within the realm of the simulated mono-enriched second-generation stars by
\citet{hartwig2018}.

From the hypothesis that \rave\ is a second-generation star\footnote{If the
assumption that \rave\ is a second-generation star is valid, then the presence
of the heavy-elements Sr and Ba in its atmosphere indicate that at least one
neutron-capture event must be accounted for in some of the first stars
\citep{roederer2014b,banerjee2018}.}, it is possible to further investigate the
characteristics of its massive stellar parent. For this, we have used the set of
theoretical nucleosynthesis yields (\texttt{znuc2012.S4}) from
\citet{heger2010}\footnote{\href{http://starfit.org}{http://starfit.org}.}, which
model the explosion of 16,800 metal-free stars with masses from 10 to
100\,$\Msun$ and explosion energies from $0.3 \times 10^{51}$\,erg to $10 \times
10^{51}$\,erg. To compare the chemical abundance pattern of \rave\ with the
theoretical values, we followed the same procedure first described in
\citet{roederer2016}, generating 10$^4$ sets of abundances by resampling the
\eps{X} and $\sigma$ values from Table~\ref{abund}, assuming gaussian
distributions. 

The results of this exercise, shown in the lower panel of Figure~\ref{acfeh} 
strongly imply ($\sim 71\%$ of the simulations)
a suitable stellar progenitor for \rave\ in the $29.5-30.0\,\Msun$ range with
an explosion energy of $10 \times 10^{51}$\,erg. In particular, the
$29.5\,\Msun$ model (black solid line) is able to reproduce the low \abund{C}{H} of
\rave\ while still providing reasonably good fits for the other elements. 
Even though the $11.3\,\Msun$ model provides the best fit in 19\% of the
simulations, its carbon abundance is consistently higher than the \rave\
detection.
The range of masses found for the progenitors of
carbon-enhanced UMP stars in \citet{placco2016b}, $29.5-30.0\,\Msun$, is
similar to the ones found here for a much lower carbon abundance. However, the
explosion energies found for the \citet{placco2016b} sample are lower by a
factor of $\sim$15-30, suggesting that this may be one of the drivers for the
distinct chemical signatures found in UMP stars.
It is also worth noting that the best fit models tend to produce lower
amounts of silicon when compared to \rave, in contrast to the lower-energy
models that better reproduce the observed Si abundance. This reinforces the need
for observing additional UMP stars, in particular with low carbon abundances.

\section{Conclusions and Future Work}
\label{final}

We have presented the first spectroscopic follow-up study of the UMP star \rave.
This star was first identified from its narrow-band \splus\ photometry.
High-resolution spectroscopy revealed a unique chemical abundance pattern, with
the lowest carbon abundance ever measured for an UMP star. Comparison with
theoretical models suggest that \rave\ is a second generation star formed in a
gas cloud polluted by the byproducts of the evolution of a progenitor in the
$\sim 30\,\Msun$ range with an explosion energy of $10 \times 10^{51}$\,erg.
Additional UMP stars identified from \splus\ photometry will greatly improve our
understanding of Pop III stars and enable the possibility of finding a
metal-free low-mass star still living in our Galaxy today.

\vspace{0.5cm}

\acknowledgments

\textcolor{white}{-}

The authors would like to thank the anonymous referee and the members of the S-PLUS community who 
provided insightful comments on the manuscript.
The work of V.M.P. is supported by NOIRLab, which is managed by the
Association of Universities for Research in Astronomy (AURA) under a
cooperative agreement with the National Science Foundation.
I.U.R. acknowledges financial support from grants AST 16-13536, AST-1815403,
and PHY 14-30152 (Physics Frontier Center/JINA-CEE) awarded by the NSF.
Y.S.L. acknowledges support from the National Research Foundation (NRF) of
Korea grant funded by the Ministry of Science and ICT (NRF-2018R1A2B6003961
and NRF-2021R1A2C1008679).
F.R.H. thanks FAPESP for the financial support through the grant 2018/21661-9.
H.D.P. thanks FAPESP for the financial support through the grant 2018/21250-9.
The S-PLUS project, including the T80-South robotic telescope and the S-PLUS
scientific survey, was founded as a partnership between the Funda\c{c}\~{a}o de
Amparo \`{a} Pesquisa do Estado de S\~{a}o Paulo (FAPESP), the Observat\'{o}rio
Nacional (ON), the Federal University of Sergipe (UFS), and the Federal
University of Santa Catarina (UFSC), with important financial and practical
contributions from other collaborating institutes in Brazil, Chile (Universidad
de La Serena), and Spain (Centro de Estudios de F\'{\i}sica del Cosmos de
Arag\'{o}n, CEFCA). We further acknowledge financial support from the São Paulo
Research Foundation (FAPESP), the Brazilian National Research Council (CNPq),
the Coordination for the Improvement of Higher Education Personnel (CAPES), the
Carlos Chagas Filho Rio de Janeiro State Research Foundation (FAPERJ), and the
Brazilian Innovation Agency (FINEP).
The members of the S-PLUS collaboration are grateful for the contributions from
CTIO staff in helping in the construction, commissioning and maintenance of the
T80-South telescope and camera. We are also indebted to Rene Laporte, INPE, and
Keith Taylor for their important contributions to the project. From CEFCA, we
thank Antonio Mar\'{i}n-Franch for his invaluable contributions in the early
phases of the project, David Crist{\'o}bal-Hornillos and his team for their
help with the installation of the data reduction package \textsc{jype} version
0.9.9, C\'{e}sar \'{I}\~{n}iguez for providing 2D measurements of the filter
transmissions, and all other staff members for their support with various
aspects of the project.
This research uses services or data provided by the Astro Data Lab at NSF’s
NOIRLab. NOIRLab is operated by the Association of Universities for Research in
Astronomy (AURA), Inc. under a cooperative agreement with the National Science
Foundation.
This research was made possible through the use of the AAVSO Photometric
All-Sky Survey (APASS), funded by the Robert Martin Ayers Sciences Fund.
This work has made use of data from the European Space Agency (ESA) mission
{\it Gaia} (\url{https://www.cosmos.esa.int/gaia}), processed by the {\it Gaia}
Data Processing and Analysis Consortium (DPAC,
\url{https://www.cosmos.esa.int/web/gaia/dpac/consortium}). Funding for the
DPAC has been provided by national institutions, in particular the institutions
participating in the {\it Gaia} Multilateral Agreement.

\software{
{\texttt{awk}}\,\citep{awk}, 
{\texttt{bokeh}}\,\citep{bokeh}, 
{\texttt{CarPy}}\,\citep{kelson2003}, 
{\texttt{gnuplot}}\,\citep{gnuplot}, 
{\texttt{IRAF}}\,\citep{tody1986,tody1993}, 
{\texttt{linemake}}\,\citep{placco2021}, 
{\texttt{MOOG}}\,\citep{sneden1973}, 
{\texttt{n-SSPP}}\,\citep{beers2014,beers2017}, 
{\texttt{sed}}\,\citep{sed}.
}

\bibliographystyle{aasjournal}
\bibliography{placco.bbl}

\end{document}